\documentstyle[prd,preprint,aps]{revtex}
\begin{document}
\draft
\title{Production of $W^{\mp}$ with an anomalous magnetic moment via the
collision of an ultrahigh-energy (anti)neutrino on a target nucleon}

\author{R. M. Garc\'{\i}a-Hidalgo and A. Rosado}
\address{Instituto de F\'{\i}sica , Universidad Aut\'onoma de Puebla.\\
Apartado Postal J-48, Colonia San Manuel, Puebla, Puebla 72570, M\'exico.}

\date{\today}
\maketitle
\begin{abstract}
We discuss the production of $W^{\mp}$ bosons in deep inelastic processes
$(\bar{\nu}_l)\nu_l + {\cal N} \rightarrow l^{\pm} + W^{\mp} + X$
(${\cal N}$: $P$ proton, $N$ neutron), in the context of an electroweak
model, in which the vector boson self interactions may be different from
those prescribed by the electroweak standard model. We present results which
show the strong dependence of the cross section on the anomalous magnetic
dipole moment $\kappa$ of the $W^{\pm}$. We show that even small deviations
from the standard model value of $\kappa$ $(\kappa =1)$ could imply
large deviations in the cross section rates of $W^{\mp}$ production
through the collision of an ultrahigh energy (anti)neutrino on a target
nucleon. However, the enhancement of the cross section rates is not
large enough to be detectable.
\end{abstract}
\narrowtext

\section{Introduction}

One of the most important features of the standard model of the electroweak
interactions \cite{stanmod} is the non-Abelian gauge nature of the vector
bosons. In the present paper, we discuss this point, namely the structure of
the self couplings of the electroweak vector bosons. In order to do it, we
analyze $W^{\mp}$ production in deep inelastic (anti)neutrino-nucleon
scattering by using the parton model \cite{partmod}, in the context of an
electroweak model with non-standard vector boson self interactions. Such a
model was proposed by M. Kuroda {\it et al.} (KMSS model) \cite{Maalampi}. In
this model the trilinear vector boson coupling constants depend on only one
free parameter, $\kappa$, the anomalous magnetic dipole moment of the
$W^{\mp}$ \cite {Neufeld}. The diagrams which contribute to the $W^{\mp}$
boson production cross section, at the quark level in the lowest order in
$\alpha$, contain only three vector boson self interactions, hence the boson
production rates depend only on $\kappa$. At the present time the best limits
on $\kappa$ are $0.76 < \kappa < 1.36$ from a recent analysis of the L3
Collaboration at LEP \cite{L3}.

In a previous work \cite{garcia}, we have discussed the heavy boson
production via the deep inelastic $\nu_l {\cal N}$ in the
context of the standard model. We pointed out there that the process
$\nu_l + {\cal N} \rightarrow l^- + W^+ + X$ (${\cal N}$: $P$ proton, $N$
neutron) is the only one which gets contribution from photon-exchange
diagrams and that its total cross section reaches significant values for
ultrahigh-energy (UHE) neutrinos\cite{uhe-neut} colliding on a target
nucleon. Therefore, we restrict ourselves our study in this paper to the
mentioned process.\footnote {the results for $\bar{\nu}_l {\cal N}$-scattering
can be obtained from those of $\nu_l {\cal N}$-scattering by the replacement
$W^+ \rightarrow W^-$, $l^- \rightarrow l^+$ and $u$-type quarks
$\leftrightarrow$ $d$-type quarks}

In this work, we calculate in the context of the standard model separately
the contribution to the process $\nu_l + {\cal N} \rightarrow l^- + W^+ + X$
from the two different mechanisms which contribute at the lowest order in
$\alpha$, keeping only photon exchange diagrams:
production at the leptonic vertex and through the boson self interaction. We
show explicitly the compensation via destructive interference inherent to the
standard model as a non-Abelian gauge theory. We find that this compensation
reaches up to four orders of magnitude, for ultrahigh-energy neutrinos with
energy $E_{\nu_l} \approx 10^{21} \, eV$ colliding on a target
nucleon. Hence, one could expect that even small deviations from the coupling
structure of the standard model, like an anomalous dipole magnetic moment
different from 1, would lead to observable effects in the predictions for
charged boson production in $\nu_l {\cal N}$ collisions.

This paper is organized as follows. In section II, we take the formulae given
in [6] for heavy boson production in $\nu_l {\cal N}$-scattering and extended
them for the case of $W^+$ boson production via the process
$\nu_l + {\cal N} \rightarrow l^- + W^+ + X$, in the frame of an electroweak
model, in which the trilinear vector boson coupling constants may deviate
through an anomalous magnetic dipole moment term from those given by the
standard model. In section III, we describe briefly the non-standard
electroweak model, which we will take for our study. In section IV, we
present and discuss our results for the total cross section of deep inelastic
process $\nu_l + {\cal N} \rightarrow l^- + W^+ + X$,
which is the only one that gets contribution from $\gamma$-exchange diagrams.
Finally, in section V, we summarize our conclusions.

\section{The differential cross section for non-standard $W^+$ production in
$\nu {\cal N}$ scattering}

We have discussed in [6] the heavy boson production via the deep
inelastic $\nu_l {\cal N}$ collisions in the context of the standard model.
We showed there that the process
\begin{equation}
\nu_l + {\cal N} \rightarrow l^- + W^+ + X\\
\end{equation}
is the only one which gets contribution from
photon-exchange diagrams and that its total cross section reaches significant
values for UHE neutrinos colliding on a target nucleon. Therefore, we restrict
ourselves our study in this paper to the mentioned reaction. In fig. 1 are
depicted the diagrams which contribute at the lowest order in $\alpha$ at the
quark parton level to the cross section of process (1).

We have presented in detail the kinematics of heavy boson production in deep
inelastic $\nu_l {\cal N} $ scattering in [6] and we pointed out there how to
take care for the different ways in which the bosons are arranged in the
non-Abelian couplings diagrams. Now we will extend the formulae given there
for the process (1) in the case in which the $W^+$ boson may have an
anomalous magnetic moment $k$ different from that predicted by the standard
model. We will assume for the trilinear coupling constants of the vector
bosons the following general form:
\begin{eqnarray}
\gamma_{\mu}(p^{0})W^{+}_{\nu}(p^{+})W^{-}_{\rho}(p^{-}) &\Rightarrow& i
g_{\gamma W^+W^-} \{ g_{\mu \nu} (\kappa p^{0}-p^{+})_{\rho} +g_{\nu \rho}
(p^{+}-p^{-})_{\mu} +  g_{\mu \rho} (p^{-} - \kappa p^{0})_{\nu} \},\\
Z_{\mu}(p^{0})W^{+}_{\nu}(p^{+})W^{-}_{\rho}(p^{-}) &\Rightarrow& i
g_{ZW^+W^-} \{ g_{\mu \nu} (\kappa _{Z} p^{0}-p^{+})_{\rho} +g_{\nu \rho}
(p^{+}-p^{-})_{\mu} +  g_{\mu \rho} (p^{-} - \kappa _{Z} p^{0})_{\nu}\}.\nonumber
\end{eqnarray}
The standard model self interactions are obtained by taking $\kappa =1$, 
$\kappa _{Z}=1$, $g_{\gamma W^+W^-}=e$ and $g_{ZW^+W^-}=e(cos \theta _{W} / 
sin \theta _{W})$ ($\theta _{W}$ being the electroweak mixing angle).

In order to calculate $W^+$ boson production through process (1) with the
non-standard non-Abelian couplings given in (2), we may take the formulas
given in section III. of [6], provided the following changes and extensions
are performed

\noindent 1) The index $r$ runs now from 1 to 4.

\noindent 2)  ${\cal F}_i^{\mu\nu}$, $f^{i,l}_{P_q,P_l}$, $f^{i,h}_{P_q,P_l}$
for $i=\{1,2,3\}$, and  $C^{l,i}_{P_q,P_l}$, $C^{l,f}_{P_q,P_l}$, 
$C^{h,i}_{P_q,P_l}$, $C^{h,f}_{P_q,P_l}$ remain as they are defined in [6].

\noindent 3) ${\cal F}_4^{\mu\nu}=\varepsilon^\mu k^\nu+k^\mu\varepsilon^\nu$ 

\noindent 4) $f^{4,l}_{P_q,P_l}=0, \; f^{4,h}_{P_q,P_l}=0$.

\noindent 5) $f^{1,n}_{P_q,P_l}=2\varepsilon (p-p') C^{n,a}_{P_q,P_l},
\; f^{2,n}_{P_q,P_l}=2 C^{n,b}_{P_q,P_l},
\; f^{3,n}_{P_q,P_l}=0,
\; f^{4,n}_{P_q,P_l}=2C^{n,c}_{P_q,P_l},$

\noindent where
\begin{eqnarray}
C^{n,x}_{L,L}&=&\sum_{B'=\gamma,Z} L^{W}_{\nu_l l'} L^{B'}_{q'q} C_{x}^{B'} 
g_{B'WW}/[(Q^2+M^2_{W})(Q'{}^2+M^2_{B'})],
\end{eqnarray}
with $x=a,b,c$. For other polarizations $L$ has to be replaced by $R$ in an
appropriate way. The values for $C_{x}^{\gamma}$ and $C_{x}^{Z}$ are given
in Table 1. We present in the Appendix the explicit expressions for the
quantities $T^{r,r'}$ and we perform also there the  summation over the
polarizations of the produced boson.

\begin{center}
\begin{tabular}{|c|c|c|c|}
\hline
\cline{2-4}
{Process}&$\hspace{1.2cm}C_{a}^{\gamma}\hspace{1.2cm}$&$\hspace{1.2cm}
C_{b}^{\gamma}\hspace{1.2cm}$&$\hspace{1.2cm}C_{c}^{\gamma}\hspace{1.2cm}$\\
\hline
$\;\; \nu_l {\cal N}\rightarrow l^- W^+ X \;\;$& $1+\kappa$&$2+(\kappa-1)/2$ 
&$(\kappa-1)/2$\\             
\hline
\end{tabular}
\end{center}

\noindent {\it Table 1. Expressions for $C_{x}^{\gamma}$ ($x=a,b,c.$) as a
function of $\kappa$ for the process (1). The expressions for $C_{x}^{Z}$ are
obtained changing $\kappa \rightarrow \kappa _{Z}$ from those given for
$C_{x}^{\gamma}$.}

\bigskip

To end this section, we want to mention the following. We have already
pointed out in [6] that, in contrast to deep inelastic lepton
nucleon scattering, the choice of the scale parameter $\tilde{Q^2}$ is not
unambiguos in the case of heavy boson production. In this work, we are going
to use the same prescription which we have used in [6].

\section{The KMSS model}

The KMSS model is an electroweak model
proposed by M. Kuroda {\it et al.} (KMSS model) \cite{Maalampi}, in which the 
bosons may have self interactions different from those prescribed by the 
standard model \cite{stanmod}. The Lagrangian of the KMSS model contains only 
dimension-four terms. P or C violation of the electromagnetic vector boson 
interactions are not allowed. In this model, the trilinear vector boson 
coupling constants can be written in the form given in (2) with

\begin{eqnarray}
\kappa_{Z} &=&e (\kappa \; tan \theta _{W} - \frac{\hat{g}}
{cos \theta _{W}})/
(e \; tan \theta _{W} - \frac{\hat{g}}{cos \theta _{W}}),\nonumber\\
g_{WW \gamma} &=& e, \\
g_{WWZ} &=&-e \; tan\theta _{W} + \frac{\hat{g}}{cos \theta _{W}},\nonumber 
\end{eqnarray}

\noindent where $\theta _{W}$ stands for the electroweak mixing angle.
$\hat{g}$ and $\kappa$ are free parameters, $\kappa$ being the anomalous
magnetic dipole moment of the $W^{\pm}$. H. Neufeld, J. D. Stroughair and D.
Schildknecht \cite{Neufeld} have considered the vector boson loop corrections
to the $\rho$ parameter in the context of the KMSS model and concluded that
in order to get good agreement with the experimental measurements:
$\rho=1 \pm 0.05$ \cite{Amaldi} the relation 

\begin{equation}
\hat{g} sin \theta _{W} =e \kappa,
\end{equation}

\noindent has to be fulfilled in very good approximation, in order to avoid
large deviation from $\rho=1$. Hence, we can reduce the set of relations
given in (4) depending on two free parameters to a set of relations
depending on only one parameter, namely $\kappa$, the $W^{\pm}$ anomalous
magnetic moment:

\begin{eqnarray}
\kappa_{Z} &=& \frac{\kappa \hspace*{1mm} cos^{2} \theta _{W}}{\kappa -
sin^{2} \theta _{W} },\nonumber\\
g_{WW \gamma} &=& e, \\
g_{WWZ} &=&e\frac{\kappa - sin^{2} \theta _{W}}{sin \theta _{W}
cos \theta _{W}}.\nonumber
\end{eqnarray}

\section{Dependence on $\kappa$ of the non-standard $W^+$ production rates}

The numerical results given in this section are obtained taking
$M_W=80.4$ $GeV$ and $M_Z=91.2$ $GeV$ for the masses of the neutral and
charged bosons (hence $sin^2 {\theta_W}=0.223$) \cite{partdata}. We leave out
in our computations of the total cross section the contribution from heavy
boson exchange diagrams. We give results for the case of unpolarized deep
inelastic $\nu_l {\cal N}$ collisions with an neutrino energy in the range
$10^{14} \, eV \leq E_{\nu_l} \leq 10^{21} \, eV$ and the nucleon at rest
($E_{\cal N}=m_{\cal N}$). We take cuts of $4$ $GeV^2$, $4$ $GeV^2$ and
$10$ $GeV^2$ for $Q^2$, ${Q'}^2$ and the invariant hadronic mass square $W$, 
respectively. These cuts are suited for the parton distribution functions 
reported by J. Pumplin {\it et al.} \cite{pumplin}, which we use in our
performances.

First, we calculate in the context of the standard model separately the
contribution to the process $\nu_l + P \rightarrow l^- + W^+ + X$ from the
two different mechanisms which contribute at the lowest order in $\alpha$
(keeping only photon exchange diagrams):
production at the leptonic vertex and through the boson self interaction.
We show in Table 2, from the first to the third column our results for
these two mechanisms (leptonic, non-Abelian) and for the total contribution.
We see in this Table the compensation via destructive interference inherent
to the standard model as a non-Abelian gauge theory. We can observe that this
compensation reaches up to four orders of magnitude. Therefore, one could
expect that even small deviations from the
coupling structure of the standard model, like an anomalous dipole magnetic
moment different from 1, would lead to observable effects in the predictions
for charged boson production in $\nu_l {\cal N}$ collisions.

Now, in order to investigate the dependence
of the $W^+$ production, via deep inelastic $\nu_l {\cal N}$-scattering, on
the $W^+$ anomalous magnetic dipole moment.
We give results for the cross section rates of the
process $\nu_l + {\cal N} \rightarrow l^- + W^+ + X$, making use of the
electroweak model proposed by M. Kuroda {\it et al.} (KMSS model)
\cite{Maalampi}. We have seen in the
previous section, that in this model the trilinear vector boson coupling
constants depend on only one free parameter, $\kappa$, the anomalous magnetic
dipole moment of the $W^{\mp}$. Further, the diagrams
which contribute to the $W^+$ boson production cross section, at the quark
level in the lowest order in $\alpha$, contain only three vector boson self
interactions, therefore the boson production rates depend only on $\kappa$.
Table 3 contains from the first to the fifth column our results for the
total cross section at the lowest order in $\alpha$ of the total cross section
of process $\nu_l + P \rightarrow l^- + W^+ + X$, taking different values for
$\kappa$. We take the values of $\kappa=$ 0.76, 0.88, 1, 1.18, 1.36 which do
not deviate too much from the SM value of $\kappa=1$ and which are allowed
according to the experimental range $0.76 <\kappa < 1.36$, reported by the L3
Collaboration \cite{L3}. We present in Table 4, the same as in Table 3, but
for the process $\nu_l + N \rightarrow l^- + W^+ + X$. Finally, we display in
Table 5, from the first to the fourth column our results for the ratio
$\sigma(\nu_l {\cal N} \rightarrow l^- W^+X)/ \sigma_{SM} (\nu_l {\cal N}
\rightarrow l^- W^+X)$. From the results given in Tables 3, 4 and 5, one can
see that the standard model prediction for the cross section is enhanced in a
84\% when we take $\kappa=1.36$ for $E_{\nu_l} = 10^{21} \, eV$. However,
this enhanced cross section rate remains unobsevable at the neutrino
telescopes \cite{telescopes}.

\begin{center}
\begin{tabular}{|c|c|c|c|}
\hline
        & \multicolumn{3}{c|}{$\sigma(\nu_l P \rightarrow l^- W^+X)$ in $cm^2$}\\
\cline{2-4}
{$E_{\nu}$ in $eV$}&$\sigma_{leptonic}$&$\sigma_{non-Abelian}$&$\sigma_{total}$\\
\hline
\hline
$10^{14}$&$\;2.45\times 10^{-37}\;$&$\;2.26\times 10^{-37}\;$&$\;4.01\times
10^{-38}\;$\\
$10^{15}$&$\;2.23\times 10^{-35}\;$&$\;2.25\times 10^{-35}\;$&$\;9.04\times
10^{-37}\;$\\
$10^{16}$&$\;3.55\times 10^{-34}\;$&$\;3.56\times 10^{-34}\;$&$\;5.02\times
10^{-36}\;$\\
$10^{17}$&$\;3.76\times 10^{-33}\;$&$\;3.77\times 10^{-33}\;$&$\;2.29\times
10^{-35}\;$\\
$10^{18}$&$\;3.80\times 10^{-32}\;$&$\;3.81\times 10^{-32}\;$&$\;1.28\times
10^{-34}\;$\\
$10^{19}$&$\;3.81\times 10^{-31}\;$&$\;3.81\times 10^{-31}\;$&$\;5.27\times
10^{-34}\;$\\
$10^{20}$&$\;3.81\times 10^{-30}\;$&$\;3.81\times 10^{-30}\;$&$\;1.23\times
10^{-33}\;$\\
$10^{21}$&$\;3.81\times 10^{-29}\;$&$\;3.81\times 10^{-29}\;$&$\;2.03\times
10^{-33}\;$\\
\hline
\end{tabular}
\end{center}
\noindent {\it Table 2. Leptonic, non-Abelian, and total contribution to the
cross section as a function of $E_{\nu}$ ($E_P=m_P$).}

\begin{center}
\begin{tabular}{|c|c|c|c|c|c|}
\hline
        & \multicolumn{5}{c|}{$\sigma(\nu_l P \rightarrow l^- W^+X)$ in $cm^2$}\\
\cline{2-6}
{$E_{\nu}$ in $eV$}&$\kappa=0.76$&$\kappa=0.88$&$\kappa=1$$\hspace*{2mm}\mbox{(S.M.)}$&$\kappa=1.18$&$\kappa=1.36$\\
\hline
\hline
$10^{14}$&$\;3.82\times 10^{-38}\;$&$\;3.91\times 10^{-38}\;$&$\;4.01\times
10^{-38}\;$&$\;4.16\times 10^{-38}\;$&$\;4.32\times 10^{-38}\;$\\
$10^{15}$&$\;8.22\times 10^{-37}\;$&$\;8.62\times 10^{-37}\;$&$\;9.04\times
10^{-37}\;$&$\;9.73\times 10^{-37}\;$&$\;1.05\times 10^{-36}\;$\\
$10^{16}$&$\;4.24\times 10^{-36}\;$&$\;4.61\times 10^{-36}\;$&$\;5.02\times
10^{-36}\;$&$\;5.71\times 10^{-36}\;$&$\;6.48\times 10^{-36}\;$\\
$10^{17}$&$\;1.70\times 10^{-35}\;$&$\;1.98\times 10^{-35}\;$&$\;2.29\times
10^{-35}\;$&$\;2.83\times 10^{-35}\;$&$\;3.44\times 10^{-35}\;$\\
$10^{18}$&$\;8.26\times 10^{-35}\;$&$\;1.04\times 10^{-34}\;$&$\;1.28\times
10^{-34}\;$&$\;1.70\times 10^{-34}\;$&$\;2.20\times 10^{-34}\;$\\
$10^{19}$&$\;3.16\times 10^{-34}\;$&$\;4.14\times 10^{-34}\;$&$\;5.27\times
10^{-34}\;$&$\;7.23\times 10^{-34}\;$&$\;9.51\times 10^{-34}\;$\\
$10^{20}$&$\;7.27\times 10^{-34}\;$&$\;9.62\times 10^{-34}\;$&$\;1.23\times
10^{-33}\;$&$\;1.70\times 10^{-33}\;$&$\;2.25\times 10^{-33}\;$\\
$10^{21}$&$\;1.19\times 10^{-33}\;$&$\;1.59\times 10^{-33}\;$&$\;2.03\times
10^{-33}\;$&$\;2.82\times 10^{-33}\;$&$\;3.73\times 10^{-33}\;$\\
\hline
\end{tabular}
\end{center}
\noindent {\it Table 3. Contribution to the total cross section as a function
of $\kappa$ and $E_{\nu}$ ($E_P=m_P$).}

\begin{center}
\begin{tabular}{|c|c|c|c|c|c|}
\hline
        & \multicolumn{5}{c|}{$\sigma(\nu_l N \rightarrow l^- W^+X)$ in $cm^2$}\\
\cline{2-6}
{$E_{\nu}$ in $eV$}&$\kappa=0.76$&$\kappa=0.88$&$\kappa=1$$\hspace*{2mm}\mbox{(S.M.)}$&$\kappa=1.18$&$\kappa=1.36$\\
\hline
\hline
$10^{14}$&$\;2.87\times 10^{-38}\;$&$\;2.92\times 10^{-38}\;$&$\;3.01\times
10^{-38}\;$&$\;3.12\times 10^{-38}\;$&$\;3.25\times 10^{-38}$\\
$10^{15}$&$\;6.66\times 10^{-37}\;$&$\;6.97\times 10^{-37}\;$&$\;7.31\times
10^{-37}\;$&$\;7.87\times 10^{-37}\;$&$\;8.45\times 10^{-37}\;$\\
$10^{16}$&$\;3.44\times 10^{-36}\;$&$\;3.75\times 10^{-36}\;$&$\;4.07\times
10^{-36}\;$&$\;4.63\times 10^{-36}\;$&$\;5.23\times 10^{-36}\;$\\
$10^{17}$&$\;1.34\times 10^{-35}\;$&$\;1.56\times 10^{-35}\;$&$\;1.81\times
10^{-35}\;$&$\;2.24\times 10^{-35}\;$&$\;2.71\times 10^{-35}\;$\\
$10^{18}$&$\;6.32\times 10^{-35}\;$&$\;7.95\times 10^{-35}\;$&$\;9.77\times
10^{-35}\;$&$\;1.30\times 10^{-34}\;$&$\;1.68\times 10^{-34}\;$\\
$10^{19}$&$\;2.38\times 10^{-34}\;$&$\;3.12\times 10^{-34}\;$&$\;3.95\times
10^{-34}\;$&$\;5.42\times 10^{-34}\;$&$\;7.12\times 10^{-34}\;$\\
$10^{20}$&$\;5.46\times 10^{-34}\;$&$\;7.21\times 10^{-34}\;$&$\;9.23\times
10^{-34}\;$&$\;1.27\times 10^{-33}\;$&$\;1.69\times 10^{-33}\;$\\
$10^{21}$&$\;8.97\times 10^{-34}\;$&$\;1.19\times 10^{-33}\;$&$\;1.52\times
10^{-33}\;$&$\;2.11\times 10^{-33}\;$&$\;2.79\times 10^{-33}\;$\\
\hline
\end{tabular}
\end{center}
\noindent {\it Table 4. Contribution to the total cross section as a function
of $\kappa$ and $E_{\nu}$ ($E_N=m_N$).}

\begin{center}
\begin{tabular}{|c|c|c|c|c|}
\hline
        & \multicolumn{4}{c|}{$\;\sigma(\nu_l {\cal N} \rightarrow l^- W^+X)/
        \sigma_{SM}(\nu_l {\cal N} \rightarrow l^- W^+X)\;$}\\
\cline{2-5}
{$\;E_{\nu}\;$ in $eV$}&$\;\kappa=0.76\;$&$\;\kappa=0.88\;$&$\;\kappa=1.18\;
$&$\;\kappa=1.36\;$\\
\hline
\hline
$10^{14}$&$\;0.95\;$&$\;0.97\;$&$\;1.04\;$&$\;1.08\;$\\
$10^{15}$&$\;0.91\;$&$\;0.95\;$&$\;1.08\;$&$\;1.16\;$\\
$10^{16}$&$\;0.84\;$&$\;0.92\;$&$\;1.14\;$&$\;1.29\;$\\
$10^{17}$&$\;0.74\;$&$\;0.86\;$&$\;1.24\;$&$\;1.50\;$\\
$10^{18}$&$\;0.65\;$&$\;0.81\;$&$\;1.33\;$&$\;1.72\;$\\
$10^{19}$&$\;0.60\;$&$\;0.79\;$&$\;1.37\;$&$\;1.80\;$\\
$10^{20}$&$\;0.59\;$&$\;0.78\;$&$\;1.38\;$&$\;1.83\;$\\
$10^{21}$&$\;0.59\;$&$\;0.78\;$&$\;1.39\;$&$\;1.84\;$\\
\hline
\end{tabular}
\end{center}
\noindent {\it Table 5. $\sigma(\nu_l {\cal N} \rightarrow l^- W^+X)/ \sigma_{SM}
(\nu_l {\cal N} \rightarrow l^- W^+X)$ as a function of $\kappa$ and $E_{\nu}$
($E_{\cal N}=m_{\cal N}$).}

\section{Conclusions}

We have analyzed the effects of a non-standard anomalous magnetic dipole
moment $\kappa$ of the $W^{\pm}$ in the charged boson production in
unpolarized deep inelastic $\nu_l {\cal N}$($\bar{\nu}_l {\cal N}$)-scattering
for UHE neutrinos colliding on a target nucleon, using the electroweak model
proposed by M. Kuroda {\it et al.} (KMSS model), in which $\kappa$ may be
different from 1, the value predicted by the electroweak standard model. We
found, that even small deviations of $\kappa$ from its standard model value
(at the present time the best limits on $\kappa$ are $0.76 < \kappa < 1.36$
from a recent analysis of the L3 Collaboration at LEP) could lead to
large enhancement in the predicted cross section rates for $W^+(W^-)$
production via $\nu_l {\cal N}$($\bar{\nu}_l {\cal N}$) collisions.
However, such enhanced rates remain too small to be detectable
experimentally at the UHE neutrino telescopes.

\bigskip

\begin{center}
{\bf ACKNOWLEDGMENTS}
\end{center}
The author thanks the {\it Sistema Nacional de Investigadores} and
{\it CONACyT} (M\'{e}xico) for financial support.

\newpage

\appendix{}
\section{}

We give in this Appendix, taking all fermion masseless, the expressions 
for the quantities $T^{r,r'}$, which are defined as follows:                                                                                                                                  
\begin{eqnarray}
T^{r,r'}_{P_q,P_l}={\cal H}_{\mu\mu'}^{P_q}{\cal F}_r^{\mu\nu}
({\cal F}_{r'}^{\mu'\nu'})^* {\cal L}_{\nu\nu'}^{P_l},\hspace{2cm}
(r,r'=1,2,3,4),
\end{eqnarray} 
with
\begin{eqnarray*}
{\cal L}^{L,R}_{\mu\mu'}&=&2\{p_{\mu} p'_{\mu'}+ p'_{\mu} p_{\mu'}- 
g_{\mu \mu'} pp' 
\mp i\varepsilon_{\mu\mu'\rho\sigma}p^{\rho} p'{}^{\sigma} \} \\
{\cal H}^{L,R}_{\mu\mu'}&=&2\{q_{\mu} q'_{\mu'}+ q'_{\mu} q_{\mu'}-
g_{\mu \mu'} qq'
\mp i\varepsilon_{\mu\mu'\rho\sigma}q^{\rho} q'{}^{\sigma} \} \\
\hspace{7mm} {\cal F}^{\mu \nu}_1&=&g^{\mu \nu} \\
\hspace{7mm} {\cal F}^{\mu \nu}_2&=&\varepsilon^{\mu} k^{\nu} - \varepsilon^{\nu} k^{\mu} \\
\hspace{7mm} {\cal F}^{\mu \nu}_3&=&i\varepsilon^{\mu \nu \rho \sigma} \varepsilon_{\rho} k_{\sigma} \\ 
\hspace{7mm} {\cal F}^{\mu \nu}_4&=&\varepsilon^{\mu} k^{\nu} + \varepsilon^{\nu} k^{\mu}
\end{eqnarray*} 
and $P_q$, $P_l$ being the polarization of the initial quark and 
incoming neutrino, respectively. In general, these quantities are 
functions of scalar products of the momenta $p$, $q$, $p'$, $q'$ , $k$
and the polarization vector $\varepsilon$ of the produced boson.

Using the polarization sum for the massive vector boson
\begin{equation}
\sum_{\lambda} \varepsilon_{\mu} (k,\lambda) \varepsilon_{\nu} (k,\lambda)=
-g_{\mu\nu}+\displaystyle \frac{k_{\mu} k_{\nu}}{M^2_B}.
\end{equation} 
and the definitions given in (A1) for the $T^{r,r'}_{, s}$ we get
\begin{eqnarray*}
\displaystyle \sum_{\lambda}{\cal F}_2^{\mu\nu}({\cal F}_4^{\rho\sigma})^*&=
&-(g^{\mu\rho}\, k^\nu 
k^\sigma+g^{\mu\sigma}\, k^\nu k^\rho)+g^{\nu\rho}\, k^\mu k^\sigma+
g^{\nu\sigma}\, k^\mu k^\rho \nonumber\\
\displaystyle \sum_{\lambda}{\cal F}_3^{\mu\nu}({\cal F}_4^{\rho\sigma})^*&=
&-i(\varepsilon^{\mu\nu\rho\beta}k^\sigma\ k_\beta+
\varepsilon^{\mu\nu\sigma\beta}k^\rho k_\beta)\\
\displaystyle \sum_{\lambda}{\cal F}_4^{\mu\nu}({\cal F}_4^{\rho\sigma})^*&=
&-(g^{\mu\rho}\, k^\nu 
k^\sigma+g^{\mu\sigma}\, k^\nu k^\rho+g^{\nu\rho}\, k^\mu k^\sigma+
g^{\nu\sigma}\, k^\mu k^\rho) 
+4k^\mu k^\nu k^\rho k^\sigma/M^2_B. \nonumber
\end{eqnarray*}
With help of these expressions we obtain ($P_qP_l=LL,RR,LR,RL$) 
\begin{eqnarray*}
\displaystyle \sum_{\lambda}T^{24}_{P_q,P_l}&=& 16(qq'\cdot kp \cdot
kp'- pp'\cdot kq \cdot kq'), \vspace{6mm}\\
\displaystyle \sum_{\lambda} Re\:T^{34}_{LL} &=&-
\displaystyle \sum_{\lambda} Re\:T^{23}_{LR},\vspace{6mm}\\
\displaystyle \sum_{\lambda} Re\:T^{34}_{RR} &=&-
\displaystyle \sum_{\lambda} Re\:T^{23}_{RL},\vspace{6mm}\\
\displaystyle \sum_{\lambda} Re\:T^{34}_{LR} &=&-
\displaystyle \sum_{\lambda} Re\:T^{23}_{LL},\vspace{6mm}\\
\displaystyle \sum_{\lambda} Re\:T^{34}_{RL} &=&-
\displaystyle \sum_{\lambda} Re\:T^{23}_{RR},\vspace{6mm}\\
\displaystyle \sum_{\lambda}T^{44}_{P_q,P_l}&=&-\displaystyle
\sum_{\lambda}T^{22}_{P_q,P_l}-16(M^{2}_{B} \; pp'\cdot qq'-
4 pk \cdot p'k \cdot qk \cdot q'k / M^{2}_{B}).
\end{eqnarray*}
     We can express $\displaystyle \sum_{\lambda} Re\:T^{14}_{P_q,P_l}$ as follows
\begin{eqnarray*}
Re\: T^{14}_{P_q,P_l}&=&\left\{ \begin{array}{llll} &8[\varepsilon p\cdot p'q \cdot q'k
+\varepsilon p'\cdot pq'\cdot qk& \hspace{0.5in} LL,RR\\
&\hspace{0.5cm} +\varepsilon q\cdot pq'\cdot p'k+
\varepsilon q'\cdot p'q\cdot pk& \\
&\hspace{0.5cm} -qq'(\varepsilon p\cdot p'k + 
\varepsilon p'\cdot pk)& \\
&\hspace{0.5cm} -pp'(\varepsilon q\cdot q'k + 
\varepsilon q'\cdot qk)] & \vspace{3mm} \\
&8[\varepsilon p\cdot p'q' \cdot qk
+\varepsilon p'\cdot pq \cdot q'k& \hspace{0.5in} LR,RL\\
&\hspace{0.5cm} +\varepsilon q\cdot p'q'\cdot pk+
\varepsilon q'\cdot pq\cdot p'k& \\
&\hspace{0.5cm} -qq'(\varepsilon p\cdot p'k + 
\varepsilon p'\cdot pk)& \\
&\hspace{0.5cm} -pp'(\varepsilon q\cdot q'k + 
\varepsilon q'\cdot qk)] & \end{array} \right.
\end{eqnarray*}
Using (A2) and the definitions given above in (A1) for the $T^{r,r'}_{, s}$
we get
\begin{eqnarray*}
\sum_{\lambda}\varepsilon P\cdot {\cal F}^{\mu \nu}_4&=&{\cal F}^{\mu 
\nu}_4(\varepsilon\rightarrow -P + \displaystyle 
\frac{kP}{M^2_B} k)  
\end{eqnarray*}
for $P=p,q,p',q'$. Hence
\begin{eqnarray*}
\sum_{\lambda}\varepsilon P\cdot Re\: T^{14}_{P_q,P_l}&=& Re\: T^{14}_{P_q,P_l} 
(\varepsilon\rightarrow -P + \displaystyle 
\frac{kP}{M^2_B} k).  
\end{eqnarray*}
The expressions for the remaining $T^{r,r'}$ can be found in the Appendix of
[6].
\bigskip

\begin{center}
{\bf Figure Caption}
\end{center}

\noindent{\bf Fig. 1}: Feynman diagrams which contribute to
process (1): (a) boson production from the incoming
neutrino, (b) the outgoing lepton, (c) the initial (d) and final quark,
and (e) through the non-Abelian couplings ($u$ stands for $u$, $c$, $\bar d$,
$\bar s$, $\bar b$; $d$ for $d$, $s$, $b$, $\bar u$, $\bar c$).

\end{document}